\documentclass[twocolumn,showpacs,preprintnumbers,amsmath,amssymb,aps,bibnotes,superscriptaddress]{revtex4}

\usepackage{amsfonts}
\usepackage[dvips]{graphicx}
\usepackage[ps2pdf,bookmarks=true,bookmarksnumbered=true,pdfborder={0 0 1}]{hyperref}
\usepackage{color}
\usepackage{bm}

\newcommand{\matrixsymb}[1]{\mathsf{#1}}
\newcommand{\vecgrk}[1]{\boldsymbol{#1}}
\newcommand{\eqnref}[1]{Eq.~\eqref{#1}}
\newcommand{\figref}[1]{Fig.~\ref{#1}}

\newcommand{\secref}[1]{Sec.~\ref{#1}}

\newcommand{\e}[1]{\text{e}^{#1}}
\newcommand{\cmplxi}{\text{i}}

\newcommand{\tr}{\operatorname{Tr}}
\renewcommand{\vec}[1]{\mathbf{#1}}

\newcommand{\punc}[1]{\,#1}

\newcommand{\neweqnline}{\nonumber\\}
\newcommand{\diffd}{\text{d}}

\def \f{{\varphi}}
\def \w{{\omega}}

\begin{document}
\title{Dynamical instability of a spin spiral in an interacting Fermi gas as a probe
of the Stoner transition}
\author{G.J.~Conduit}
\email{gjc29@cam.ac.uk}
\affiliation{Department of Condensed Matter Physics, Weizmann Institute of Science, Rehovot, 76100, Israel}
\affiliation{Physics Department, Ben Gurion University, Beer Sheva, 84105, Israel}
\author{E.~Altman}
\affiliation{Department of Condensed Matter Physics, Weizmann Institute of Science, Rehovot, 76100, Israel}
\date{\today}

\begin{abstract}
We propose an experiment to probe ferromagnetic phenomena in an
ultracold Fermi gas, while alleviating the sensitivity to three-body
loss and competing many-body instabilities.  The system is initialized
in a small pitch spin spiral, which becomes unstable in the presence
of repulsive interactions. To linear order the exponentially growing
collective modes exhibit critical slowing down close to the Stoner
transition point. Also, to this order, the dynamics are identical on
the paramagnetic and ferromagnetic sides of the transition. However,
we show that scattering off the exponentially growing modes
qualitatively alters the collective mode structure. The critical
slowing down is eliminated and in its place a new unstable branch
develops at large wave vectors. Furthermore, long-wavelength
instabilities are quenched on the paramagnetic side of the
transition. We study the experimental observation of the
instabilities, specifically addressing the trapping geometry and how
phase-contrast imaging will reveal the emerging domain structure. These
probes of the dynamical phenomena could allow experiments to detect
the transition point and distinguish between the paramagnetic and
ferromagnetic regimes.
\end{abstract}


\maketitle

\section{Introduction}

A magnetic field tuned Feshbach resonance provides a powerful tool to
control the interaction parameters of ultracold atomic Fermi
gases~\cite{76s12}. The effective interaction between two atoms in an
s-wave scattering channel is attractive on one side of the resonance
and repulsive on the other side, with both regimes diverging upon
approaching the resonance. Over the last few years experiments
starting from the attractive side have investigated the crossover from
a Bardeen-Cooper-Schrieffer state of fermion pairs to a Bose-Einstein
condensate of tightly bound molecules~\cite{03sph08}. On the other
hand, a recent experiment has provided the first possible evidence for
a transition to an itinerant ferromagnet beyond a critical interaction
strength in the repulsive regime~\cite{Jo09}.  If confirmed, this new
realization of ferromagnetism may not only resolve long-standing
questions stemming from the solid state but also promises to open up
new arenas of ferromagnetism
research~\cite{Duine05,Coleman08,Conduit08,Zhai09,Duine10}.

When considering the repulsive side of the resonance however, it must
be noted that the repulsive Fermi gas is only a meta-stable state.
The two-body ground state in this regime is a ``Feshbach molecule'', a
bound state with negative energy.  Correspondingly, the many fermion
ground state is the molecular BEC, whereas the repulsive Fermi gas
arises only if the system is specially prepared without molecules.
Even then, atoms gradually recombine to form Feshbach molecules.  At
least in the low density limit, this occurs predominantly through a
three-body process~\cite{Petrov03}, whereas closer to the resonance the
loss may reflect competing many-body
instabilities~\cite{Eugene-private}.  An experiment as in
Ref.~\cite{Jo09} must therefore be performed inherently out of
equilibrium. To abate the fall in atom density the experiment was
performed while tuning the interaction parameter rapidly, but this
however masks the true phase
transition~\cite{Conduit10,Babadi09}. Moreover, it has been shown that
even if the atom number is kept constant by coupling the system to an
atom reservoir, the non-equilibrium conditions imposed by the
three-body loss act to change the nature of the ferromagnetic
transition through an inherently quantum mechanism~\cite{Conduit10ii}.

In this paper we propose a different strategy to investigate the
Stoner transition, which could allow investigators to circumvent the
difficulties imposed by atom loss. The idea is to study the dynamical
stability of a nearly ferromagnetic state, or more precisely a spin
spiral of small wave-vector ${\bf{Q}}$, as shown in
\figref{fig:CantedSpinWave}(b). This state has minimal three-body
losses as it is locally fully polarized.  On the other hand, the
dynamical stability of the spiral spin texture is not protected by
spin conservation because the system has zero net magnetization. To
take advantage of this new protocol it is first vital to determine
how the modes of instability change when we tune the system across the
Stoner transition. For an interaction strength tuned so that
ferromagnetism is favored, the exponentially growing unstable modes are
expected to reorient the spins, as seen in
\figref{fig:CantedSpinWave}(c), and eventually cause the system to
fragment into polarized domains. A similar collective modes structure
was observed in a bosonic ferromagnetic gas~\cite{Cherng08}.  Unlike
when ordering from the paramagnetic state~\cite{Babadi09}, the size of
these domains and the collective mode structure can be finely tuned
with the length scale of the initial spin spiral.

On approaching the Stoner transition we find critical slowing down of
the unstable modes.  However, counter to initial heuristic
expectations, the unstable modes of the helical spin state are, to
linear order, the same on the two sides of the transition. To
differentiate between the ferromagnetic and paramagnetic regimes we go
beyond the linear analysis and study the feedback effect due to the
scattering of collective excitations off the exponentially growing
modes. As the phenomena investigated are not only of conceptual
interest but could also provide a new protocol for the next generation
of experiments, we also consider the ramifications of a realistic
harmonic trapping potential and the experimental probes of the
collective modes.

\section{Forming the spin spiral}\label{sec:FormSpinSpiral}

\begin{figure}
 \centerline{\resizebox{0.9\linewidth}{!}{\includegraphics{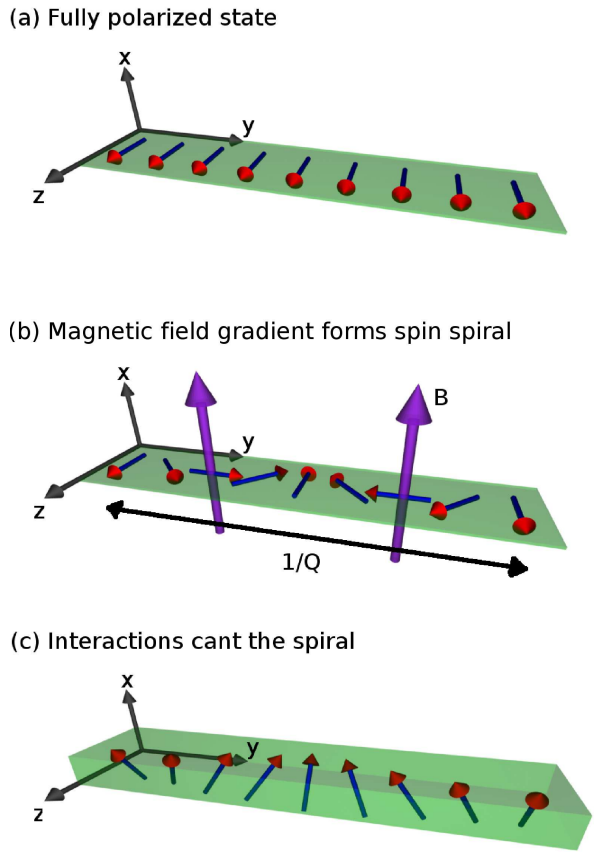}}}
 \caption{(Color online) (a) The gas is started in a fully polarized
   state and (b) a normal magnetic field $B$ (purple arrows) is
   applied to form a spin spiral of pitch $1/Q$. When the interaction
   strength is ramped upwards (c) the spins rotate into the y
   direction forming a fully polarized state (which we later show has
   a long wavelength modulation). The thickness of the green plane
   indicates the growing magnetization. In (b) and (c) the axis basis
   set co-rotates with the spin spiral.}
 \label{fig:CantedSpinWave}
\end{figure}

To form the initial spin spiral, the atomic gas is first prepared in a
fully polarized phase, say along the z spin axis shown in
\figref{fig:CantedSpinWave}(a), and a magnetic field gradient is
imposed perpendicular to the magnetization axis (e.g. ${\bf
  B}=by{\bf\hat x}$) for time $t$. Such a field can be thought of as a
gradient of the relative potential between the $\uparrow_{\text{x}}$
and $\downarrow_{\text{x}}$ spins, and leads to a (constant) relative
acceleration between these two spin components in the coherent spin
state.  The result is a spiral spin texture, as shown in
\figref{fig:CantedSpinWave}(b), with wave vector
$Q_{\text{y}}=(\mu_{\text{B}}g_{\text{J}}t/\hbar)\diffd
B_{\text{x}}/\diffd y$, where $g_{\text{J}}$ is the $g$-factor. The
twist rate is independent of the spin stiffness and the strength of
the repulsive interactions between particles. This is in close analogy
to the effect of a potential gradient placed across a superfluid,
which is essentially a XY ("phase") ferromagnet. The potential
gradient affects phase twist at a constant rate, independent of the
superfluid stiffness, which corresponds to free acceleration according
to Newton's law.

To gain further understanding into why interactions do not impact on
the dynamic formation of the initial spin spiral we can study the
system within the simple setting of the Heisenberg ferromagnet. Then,
the situation posed in \figref{fig:CantedSpinWave} is described by the
Hamiltonian $\hat{H}=-J\sum_{\langle
  ij\rangle}\hat{\vec{S}}_{i}\cdot\hat{\vec{S}}_{j}-b\sum_{i}y_{i}\hat{S}^{\text{x}}_i$,
where the first summation covers nearest neighbors on the lattice, $J$
denotes the coupling between adjacent sites, and $b$ is the magnetic
field gradient. The initial conditions are $S^{\text{x}}_i(t=0)=0$,
$S^{\text{y}}_i(t=0)=0$, and $S^{\text{z}}_i(t=0)=S_{0}$.  We then
study the evolution of the spins using
$\hbar\dot{\vec{S}}=\cmplxi[\hat{H},\vec{S}]$, finding the equations
of motion $\hbar\dot{S}_{\text{x},i}=0$, $\hbar\dot{S}_{\text{y},i}=b
y_{i}S_{\text{z},i}$, and $\hbar\dot{S}_{\text{z},i}=-b
y_{i}S_{\text{y},i}$, where the components containing $J$ cancel
exactly. Finally, we can then solve the equations of motion to yield
$S_{\text{x},i}(t)=0$, $S_{\text{y},i}(t)=S_{0}\sin(b y_{i}t/\hbar)$,
and $S_{\text{z},i}(t)=S_{0}\cos(b y_{i}t/\hbar)$. These show that the
formation of the spin spiral by the external magnetic field gradient
is independent of the interactions, $J$, between particles.

An interesting and beneficial practical implication of the above
observation is to alleviate the need to first form a spiral, and then
perform a Feshbach field quench. The spin spiral may equally well be
formed with the Feshbach field in place. Finally we note that a
magnetic field gradient has the side effect of imparting a
translational force on the gas perpendicular to the
gradient. Fortunately, in this scheme we need to impose only a
long-pitched spin spiral, which requires only a weak magnetic field
gradient with minimal side effect.

Following the preparation stage, we are ready to allow the spiral
state to evolve under the influence of the repulsive interactions
tuned by the Feshbach field, and track its evolution into the
polarized state out of the plane of the initial spiral shown in
\figref{fig:CantedSpinWave}(c).

\section{Linear spin-wave instability}\label{sec:FirstOrder}

We now investigate the instabilities of the helical spin state by
first focusing on the linearized spin fluctuations around the initial
spin spiral. The exponentially growing unstable modes will show up in
this analysis as collective excitations with imaginary frequencies.
As the dominant modes grow exponentially, non-linear processes become
important. Later in \secref{sec:NonlinearCollectiveModes} we will
study how scattering off the exponentially growing modes renormalizes
the spectrum.

To study the collective modes we start from the quantum partition
function expressed as a fermionic coherent state path integral,
$\mathcal{Z}=\tr\e{-\beta(\hat{H}-\mu\hat{N})}=\int\mathcal{D}\psi\e{-S}$,
with the corresponding action
\begin{align}
 S=\int\!\!\!\!\sum_{\sigma=\{\uparrow,\downarrow\}}\!\!\bar{\psi}_{\sigma}
 \left(\partial_{\tau}+\epsilon_{\hat{\vec{k}}}-\mu\right)\psi_{\sigma}+\!\!
 \int g\bar{\psi}_{\uparrow}\bar{\psi}_{\downarrow}\psi_{\downarrow}\psi_{\uparrow}\punc{,}
\end{align}
where $\int\equiv\int_{0}^{\beta}\diffd\tau\int\diffd\vec{r}$, the
free particle dispersion is $\epsilon_{\vec{k}}=k^{2}/2$, $\mu$ is the
chemical potential, and $g\delta^{3}(\vec{r})$ is the strength of the
s-wave repulsive contact interaction. We have also set $\hbar=m=1$.
To explore how interactions impinge on the collective mode spectrum we
affect a Hubbard-Stratonovich decoupling, which incorporates the spin
channels $\vecgrk{\phi}_{0}+\vecgrk{\phi}$
\cite{Conduit08,Conduit09}. The static component of the magnetization,
$\vecgrk{\phi}_{0}$, follows the initial helical spatial spin texture
and $\vecgrk{\phi}$ represents the growing unstable modes and
fluctuations around that stationary component. It is also convenient
to apply a gauge transformation
$\psi\mapsto\psi\e{\cmplxi\vec{Q}\cdot\vec{r}\sigma_{\text{x}}/2}$ to
enter a spatially rotating basis set with pitch vector
$Q\vec{e}_{\text{x}}/2$, which renders the initial spin texture, and
thus also the static component of the magnetization to be uniform,
$\vecgrk{\phi}_{0}=(0,0,\phi_{0})$.

After integrating out the Grassmann fields we obtain
$\mathcal{Z}=\int\mathcal{D}\vecgrk{\phi}\e{-S}$ with
the action
\begin{align}
 S\!&=\!\!\int\!g\phi^{2}_{0}\!-\!\tr\ln\left(\hat{\matrixsymb{G}}_{0}^{-1}\right)\neweqnline
 \!&+\!\!\int\!g\vecgrk{\phi}^{2}\!-\!\tr\ln\left[\matrixsymb{I}\!+
 \!\hat{\matrixsymb{G}}_{0}\left(\frac{1}{2}\sigma_{\text{x}}\vec{Q}\cdot\hat{\vec{k}}\!-\!g\vecgrk{\sigma}\cdot\vecgrk{\phi}\right)\!\right]\!\punc{,}
\label{Sphi}
\end{align}
where $\hat{\matrixsymb{G}}_{0}^{-1}=
\partial_{\tau}+\epsilon_{\hat{\vec{k}}}-\mu-g\sigma_{\text{z}}\phi_{0}=\hat{G}_{\sigma_{\text{z}}}^{-1}$
denotes the elements of the inverse Green function at the level of the
renormalized mean field.  We then expand in the perturbative
collective modes $\vecgrk{\phi}$. We assume that the initial spin
spiral has a long wavelength relative to the Fermi wave vector
$k_{\text{F}}$, so we work in the regime $Q\ll k_{\text{F}}$. Focusing
on the soft modes that are perpendicular to the saddle point field,
the dispersion satisfies $\omega\ll\mu$ and has wave vector $q\ll
k_{\text{F}}$. Following the expansion for the soft modes we get the
contribution to the action from fluctuations
\begin{align}
 &S\!=\!\!\int g(\phi_{\text{x}}^{2}+\phi_{\text{y}}^{2})\neweqnline
 &\!+\!\tr\!\left\{\!\hat{G}_{+}\!\left[\frac{\vec{Q}\!\cdot\!\vec{k}}{2}\!-\!g(\phi_{\text{x}}\!-\!\cmplxi\phi_{\text{y}})\right]\!\hat{G}_{-}\!\left[\frac{\vec{Q}\!\cdot\!\vec{k}}{2}\!-\!g(\phi_{\text{x}}\!+\!\cmplxi\phi_{\text{y}})\right]\!\right\}\neweqnline
 &\!+\!\frac{1}{2}\!\tr\!\left\{\!\hat{G}_{+}\!\left[\frac{\vec{Q}\!\cdot\!\vec{k}}{2}\!-\!g(\phi_{\text{x}}\!-\!\cmplxi\phi_{\text{y}})\right]\!\hat{G}_{-}\!\left[\frac{\vec{Q}\!\cdot\!\vec{k}}{2}\!-\!g(\phi_{\text{x}}\!+\!\cmplxi\phi_{\text{y}})\right]\!\right\}^{2}\!\!\!\!
\punc{.}
\end{align}
Before searching for the collective modes it is useful to transform
the basis set for the magnetization from Cartesian to spherical polar
coordinates:
$\phi=\phi_{0}[1+\eta](\cos\theta,\sin\theta\sin\f,\sin\theta\cos\f)$.
Note that $\theta$ is defined as the angle with respect to the
positive x axis, and $\f$ is the angle from the z axis of the
projection of ${\vec \phi}$ onto the y,z plane.  The initial state is
$\theta=\pi/2$ and $\f=0$, which ensures that fluctuations in $\f$ are
well defined. With this definition we expand in small thermal and
quantum fluctuations away from the fully polarized state that are both
angular $\f$ and $\theta$ (now redefined to be centered around
$\pi/2$), and also in the amplitude of the mode, $\eta$. The
perturbative form for the expansion correct to quadratic order is
$\vecgrk{\phi}=\phi_{0}[1+\eta](-\theta,\f,1-[\theta^{2}+\f^{2}]/2)$. Although
both Cartesian and spherical polar basis sets yield the same
collective mode structure in this linear response analysis, when in
\secref{sec:NonlinearCollectiveModes} we consider feedback corrections
to this response it will be necessary to work with the spherical polar
basis set to properly evaluate phase and amplitude fluctuations. However,
to study the system with only linear response we also expand $G_{\pm}$
in $\omega$ and $\vec{q}$ and find that the soft modes are coupled in
the action through
\begin{align}
 S\!=\!S_{0}\tr\!\left\{\!\left(
    \begin{array}{cc}
    \!\theta_{-\vec{q}}^{-\omega}&\f_{-\vec{q}}^{-\omega}\!
    \end{array}
  \right)\!\!
   \left(
    \begin{array}{cc}
     \!\chi[q^{2}-Q^{2}]&\cmplxi\omega\!\\
     \!-\cmplxi\omega&\chi q^{2}\!
    \end{array}
   \right)\!\!
   \left(
    \begin{array}{c}
     \!\theta_{\vec{q}}^{\omega}\!\\\!\f_{\vec{q}}^{\omega}\!
    \end{array}
   \right)\!\right\}\!\!
   \punc{,}
   \label{eqn:MatrixEqnForModes}
\end{align}
where $\chi=\frac{1}{2}-\frac{2^{2/3}3}{5k_{\text{F}}a}$,
$q^{2}=q_{\text{x}}^{2}+q_{\text{y}}^{2}+q_{\text{z}}^{2}$, and
$S_{0}=2^{1/3}\beta\phi_{0}k_{\text{F}}a/\pi$. The twist wave vector
$Q$ couples only to the magnetization along the initial spin
spiral. Due to their commutation relations, the conjugate modes are
coupled by the off-diagonal elements and so to extract the dispersion
we demand that the determinant of the matrix is zero. This yields the
dispersion
\begin{align}
 \omega=\pm\chi q\sqrt{q^{2}-Q^{2}}\punc{,}
 \label{eqn:ContinuumBareDispersion}
\end{align}
which is shown in \figref{fig:RenormalizedSurfaceGrowth}(a).  We first
verify the dispersion in the absence of the spin spiral. When $Q=0$
and interactions are strong so $\chi\to 1/2$ we recover the familiar
dispersion of a single particle, $\omega=q^{2}/2$. When $q<Q$ the
dispersion is imaginary, corresponding to an instability, whereas when
$q>Q$ we recover oscillating modes. Note that, as required by spin
conservation, there is no dynamical instability of the magnetization
at zero wave-vector.  Similarly growth at $q=Q$ is stunted as this
mode is initially fully polarized. The exponential growth of the order
parameter is maximal at the wave-vector $q=Q/\sqrt{2}$, and it is at
this length scale that we would expect to see ferromagnetic domains
emerge. In the experiment~\cite{Jo09} domain walls could not be
observed in a gas starting from a paramagnetic state since their size
falls below the resolution of current experimental
techniques~\cite{Babadi09}, but here their length scale can be tuned
to be experimentally observable by changing the pitch of the spin
spiral. The fully polarized phase becomes unstable at the critical
interaction strength $k_{\text{F}}a=2^{5/3}3/5\approx1.90$. At this
interaction strength $\chi=0$ so the system undergoes critical slowing
and the characteristic time of the instability diverges.

Against intuitive expectations we find the same collective mode
structure, to linear order, on either side of the transition. To gain
insight on this, it is useful to study the linear stability of a
spiral texture in the Heisenberg model on a lattice
$\hat{H}=-J\sum_{\langle
  ij\rangle}\hat{\vec{S}}_{i}\cdot\hat{\vec{S}}_{j}$. Here $\langle
ij\rangle$ restricts the summation to nearest neighbors and $J$
denotes the coupling between adjacent sites. A straightforward spin
wave analysis of the fluctuations about the spiral texture yields the
collective mode dispersion
\begin{align}
 &\omega_{\text{H}}\!=\!4JS\!\sqrt{\cos^{2}\!(Qa)\sin^{2}\!
 \left(\!\frac{q_{\text{x}}a}{2}\!\right)\!\!+\!\sin^{2}\!\left(\!\frac{q_{\text{y}}a}{2}\!\right)\!\!+
 \!\sin^{2}\!\left(\!\frac{q_{\text{z}}a}{2}\!\right)\!}\neweqnline
 &\times\!\!\sqrt{\sin^{2}\!\left(\!\frac{q_{\text{x}}a}{2}\!\right)\!\!+\!\sin^{2}\!
 \left(\!\frac{q_{\text{y}}a}{2}\!\right)\!\!+\!\sin^{2}\!\left(\!\frac{q_{\text{z}}a}{2}\!\right)\!\!
 -\!\sin^{2}\!\left(\!\frac{Qa}{2}\!\right)}\punc{,}
 \label{Disp-Heisenberg}
\end{align}
where $a$ is the lattice spacing. Note that in the limit $qa\ll1$ and
$Qa\ll1$ the collective mode dispersion \eqnref{Disp-Heisenberg}
computed in the Heisenberg model approaches the same form
\eqnref{eqn:ContinuumBareDispersion} developed in the continuum case.
Tuning the coupling from ferromagnetic to antiferromagnetic through
criticality at $J=0$ is allied with a vanishing dis[persion, as found
at the Stoner transition of the Fermi system. The anisotropy
introduced by the first root gives a preference to modes with a wave
vector in the plane of the spiral.

We note that the equivalent dynamics of the initial spin-spiral seen
for positive and negative $J$ is due to an exact symmetry of the
Heisenberg Hamiltonian.  This can be shown by a variation on the
arguments presented in Ref.~\cite{Sorensen09}, which is valid for
initial states invariant under some generalized time reversal
transformation.  Such an exact symmetry does not exist in the
itinerant fermion model of interest here. Rather the symmetry of the
collective mode dispersion around the dynamical critical point is an
emergent phenomenon, and as we shall see later in
\secref{sec:NonlinearCollectiveModes}, valid only within the linear
approximation of the dynamics.

Before proceeding to study these nonlinear effects, we revisit our
initial assumption that the three-body loss may be neglected in the
proposed setup.  To assess the validity of this assumption we shall
compare the loss rate in the spin spiral to the growth rate of the
maximally unstable mode around this spiral state. The three-body loss
rate (strictly valid for $k_{\text{F}}a\ll 1$) is
$111\bar{\epsilon}(k_{\text{F}}a)^{6}n_{\uparrow}n_{\downarrow}(n_{\uparrow}+n_{\downarrow})$~\cite{Petrov03}. To
evaluate this we note that adjacent spins in the spiral, separated by
$r_{\text{s}}=(2/9\pi)^{1/3}k_{\text{F}}^{-1}$, have an angle between
them of $Qr_{\text{s}}\ll1$. Therefore the adjacent spins give a
geometric component of
$n_{\uparrow}n_{\downarrow}=(2/9\pi)^{2/3}n^{2}(Q/k_{\text{F}})^{2}/4$. Applying
this to the experimental regime gives the loss rate
$0.12(k_{\text{F}}a)^{6}(Q/k_{\text{F}})^{2}\,\text{s}^{-1}$.

On the other hand, the growth rate of the dominant mode, according to
\eqnref{eqn:ContinuumBareDispersion}, is
$\max_{\vec{q}}\{\text{Im}[{\omega}(\vec{q})]\}=(1-2^{5/3}3/5k_{\text{F}}a)Q^{2}/4$. Interestingly
we observe that the ratio of loss to growth rate depends only on the
interaction strength and not on the spiral pitch $Q$. Comparing the
two rates we find that the loss will dominate for strong interactions
$k_{\text{F}}a\gtrsim 4.9$, though in this regime the formula for loss
rate significantly overestimates the true loss~\cite{Huckans09} and in
fact the theory should be valid for even higher interaction
strengths. However, in the experimentally accessible region including
the phase transition, $k_{\text{F}}a<2.5$, loss is more than 60 times
smaller than the dominant growth rate of the collective
modes. Therefore the new experimental protocol offers a promising way
to observe magnetism without the damaging effects of loss. In
addition, losses will dominate within a small region of width $\delta
(k_{\text{F}} a)\sim 0.03$ surrounding the point of critical slowing
down. However, we shall see in the next section that non-linear
feedback effects drive a new dynamical instability at large
wave-vectors precisely in the region of critical slowing down, and that
this instability will overcome the losses.

\section{Nonlinear collective modes}\label{sec:NonlinearCollectiveModes}

\begin{figure*}
 \centerline{\resizebox{1.0\linewidth}{!}{\includegraphics{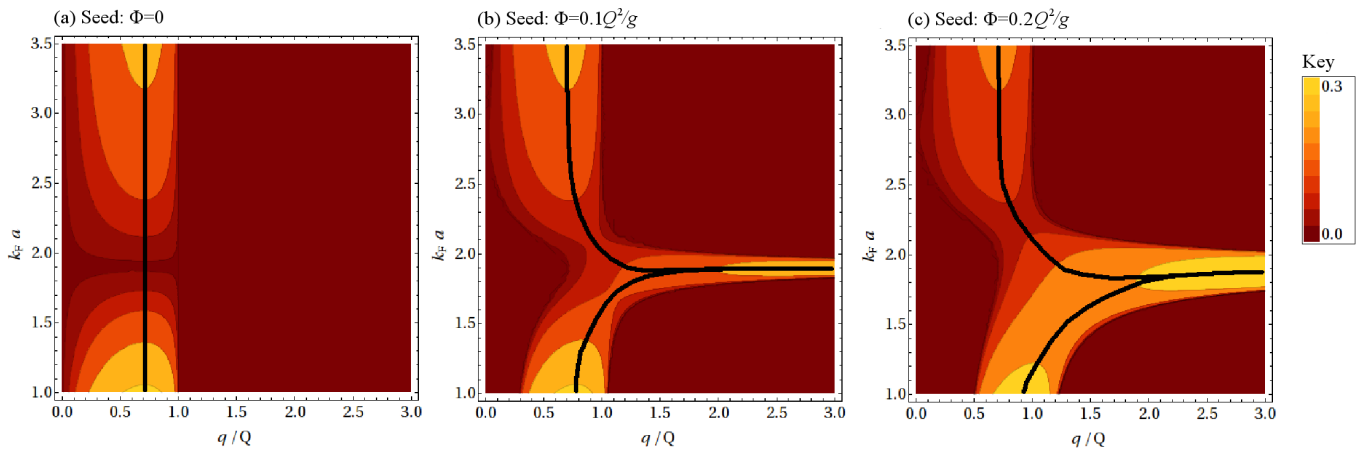}}}
 \caption{(Color online) The variation of the exponential growth rate
   of the collective mode at wave vector $\vec{q}$ with interaction
   strength $k_{\text{F}}a$. Calculated for an initial seed field of
   size (a) $\Phi=0$, (b) $\Phi=0.1Q^{2}/g$, and (c)
   $\Phi=0.2Q^{2}/g$. The fields $\{\theta,\f\}$ are evaluated at wave
   vector $\vec{q}$, and $\eta$ at wave vector $\vec{q}\pm\vec{Q}$; we
   focus on the peak contribution to the growth with
   $\vec{q}\parallel\vec{Q}$. The black trajectory highlights the
   maximum growth rate at a given interaction strength.}
 \label{fig:RenormalizedSurfaceGrowth}
\end{figure*}

In the previous section we found that the behavior of the collective
modes spectrum with interaction strength is qualitatively similar
either side of the critical interaction strength
$k_{\text{F}}a=2^{5/3}3/5$. However, for the Fermi system this is only
a feature of the linear analysis, which we now extend to include,
self-consistently, the effect of the scattering of fluctuations on the
exponentially growing modes. We will see that this mechanism also
eliminates the region of critical slowing down by generating new
unstable modes at high wave-vectors near to the critical interaction
strength.

We focus solely on the consequences of scattering off a single
dominant growth mode $\vecgrk{\Phi}_{\vec{P}}$. According to the
analysis of \secref{sec:FirstOrder}, the momentum $\vec{P}$ associated
with the dominant mode is $Q/\sqrt{2}$, though here we allow a general
wave vector that we will later determine self consistently. In
presence of the large mode $\Phi_{\vec{P}}$, we have to expand the
action \eqnref{Sphi} to cubic order to include terms which are linear
in $\Phi_{\vec{P}}$ and quadratic in the other modes
\begin{align}
 S_{\Phi}&=\int g^{3}\tr\left[G_{-}(\Phi_{\text{x}}+\cmplxi\Phi_{\text{y}})G_{+}\phi_{\text{z}}G_{+}(\phi_{\text{x}}-\cmplxi\phi_{\text{y}})\right.\neweqnline
   &-\left.G_{+}(\Phi_{\text{x}}-\cmplxi\Phi_{\text{y}})G_{-}\phi_{\text{z}}G_{-}(\phi_{\text{x}}
   +\cmplxi\phi_{\text{y}}) \right]\punc{,}
\end{align}
where $G_{\pm}=\partial_{\tau}+\epsilon_{\hat{\vec{k}}}-\mu\mp
g\phi_{0}$.  This new scattering mechanism couples the
$\{\phi_{\text{x}},\phi_{\text{y}}\}$ channels to the
$\phi_{\text{z}}$ channel, and so describes scattering out of the
original magnetization configuration and into the dominant growing
mode. The presence of a growing classical field
$\Phi(t)=\Phi(0)e^{i\Omega \tau}$ (with imaginary frequency $\Omega$),
requires a seed fluctuation in the initial spiral state. Such a seed
will be present in any realistic implementation because of random
inhomogeneity in the magnetic field and thermal fluctuations.

We again transform to the spherical polar basis set to properly
separate phase and amplitude fluctuations. Coupling to the
exponentially growing mode of wave vector $\vec{P}$ and frequency
$\Omega$ requires studying the spin susceptibility matrix expanded
out to include couplings between other modes at wave vectors $\vec{q}$
and $\vec{q}\pm\vec{P}$, and frequencies $\w$ and $\w\pm\Omega$. The
coupled action then takes the form
\begin{widetext}
\begin{align}
 S\!=\!S_{0}\tr\!\left\{\!\!\left(\!
    \begin{array}{cccc}
     \theta_{-\vec{q}}^{-\omega}&\f_{-\vec{q}}^{-\omega}&\eta_{-\vec{q}-\vec{P}}^{-\omega-\Omega}&\eta_{-\vec{q}+\vec{P}}^{-\omega+\Omega}
    \end{array}
  \!\right)\!\!
   \left(
    \begin{array}{cccc}
     \!\!\chi[q^{2}\!-\!Q^{2}]&\cmplxi\omega&\zeta&\zeta\\
     -\cmplxi\omega&\chi q^{2}&\zeta&\zeta\\
     \zeta&\zeta&\frac{3E_{\text{F}}}{2}\!\left[\!\chi\!-\!\frac{\pi(\omega+\Omega)}{\sqrt{2E_{\text{F}}}|\vec{q}+\vec{P}|}\!\right]&0\\
     \zeta&\zeta&0&\frac{3E_{\text{F}}}{2}\!\left[\!\chi\!-\!\frac{\pi(\omega-\Omega)}{\sqrt{2E_{\text{F}}}|\vec{q}-\vec{P}|}\!\right]\!\!
    \end{array}
   \right)\!\!
   \left(
    \begin{array}{c}
     \theta_{\vec{q}}^{\omega}\\\f_{\vec{q}}^{\omega}\\\eta_{\vec{q}+\vec{P}}^{\omega+\Omega}\\\eta_{\vec{q}-\vec{P}}^{\omega-\Omega}
    \end{array}
   \right)\!\!\right\}\!\!
   \punc{,}
   \label{eqn:BigMatrixEqnForModes}
\end{align}
\end{widetext}
where
$\zeta=3E_{\text{F}}\theta_{\text{s}}(1-\frac{2^{2/3}9}{5k_{\text{F}}a})/2$,
$E_{\text{F}}$ is the polarized state Fermi energy, and
$\theta_{\text{s}}$ denotes the seed field size for either $\theta$ or
$\f$. The expansion in frequencies employed for the amplitude modes,
$\eta$, applies for $\omega\ll q$, which holds for the regime of
interest that describes the coupling the soft angular modes to the
amplitude modes.  We again search for the zeros in the resultant
determinant to extract the collective modes and explore the unstable
region. Note that in the absence of the growing field, the formalism
immediately recovers the dispersion found in
\secref{sec:FirstOrder}. However, here we aim to go further and
consider the form of the modes in the presence of the dominant growing
field. To ensure that the formalism is self consistent at each
interaction strength we first compute the feedback-corrected dominant
mode before calculating the entire collective modes spectrum in the
presence of that growing mode. We also note that this formalism
respects the the spin conservation law, and so, as in the linear
analysis of \secref{sec:FirstOrder}, growth of the uniform component
of the magnetization is suppressed.

The final result for the spectrum is shown in
\figref{fig:RenormalizedSurfaceGrowth} for different values of the
seed field. First we note that for interaction strengths far from the
critical slowing, scattering on the growing mode is negligible. In
particular the dominant growth mode remains at $\sim Q/\sqrt{2}$ in
this regime. However, near to critical slowing the wave vector of the
dominant growth mode can be enhanced. This can be understood by
comparing the on and off-diagonal elements of the action matrix,
\eqnref{eqn:BigMatrixEqnForModes}. Upon nearing critical slowing, the
on-diagonal elements, being proportional to $\chi$, approach
zero. However, the off-diagonal terms, $\zeta$, that represent
scattering off the dominant growing modes are non-zero. To ensure that
the overall determinant is zero demands a non-zero value for the
on-diagonal elements which requires a large wave vector. On
approaching critical slowing this growth in the wave vector would only
be curtailed by higher order momentum terms. A detailed analysis shows
that the wave vector of the peak growth tracks the trajectory along
the critical slowing at $k_{\text{F}}a=2^{5/3}3/5$ in the phase
diagram \figref{fig:RenormalizedSurfaceGrowth}. The peak growth
  at a particular wave vector $\vec{q}$ is when that wave vector
  $\vec{q}$ is parallel to the direction of $\vec{Q}$, and so we focus
  on that contribution in \figref{fig:RenormalizedSurfaceGrowth}. On
the paramagnetic side of the resonance, $k_{\text{F}}a<2^{5/3}3/5$,
the enhanced scattering into the higher momentum sector removes long
wavelength components of the domains, whereas on the ferromagnetic
side of the resonance, $k_{\text{F}}a>2^{5/3}3/5$, the larger domains
are naturally still favored. Therefore, consideration of the feedback
of the exponentially growing field yields an additional collective
mode structure that could experimentally distinguish between the two
sides of the critical slowing interaction strength.

This picture was developed in the presence of just a single growing
exponential mode, whereas in reality these seed modes have an initial
growth rate $\sim\chi q\sqrt{Q^{2}-q^{2}}$
(\eqnref{eqn:ContinuumBareDispersion}) centered around the wave vector
$q=Q/\sqrt{2}$ corresponding to maximal growth. With reference to
\figref{fig:RenormalizedSurfaceGrowth}, except near to critical
slowing, at a given interaction strength the mode growth rate is
sharply peaked as a function of $q$ so that the presence of other less
dominant exponentially growing modes blurs the collective modes
dispersion by less than $5\%$. The relative initial size of those
modes can however have an impact on the mode spectrum. Created by
inhomogeneities in the applied magnetic field, the uncertainty can
correspond to seeing a range of growth rates, for example the range
spanning between \figref{fig:RenormalizedSurfaceGrowth}(a) and
\figref{fig:RenormalizedSurfaceGrowth}(c). Though the growth rate
around critical slowing is quantitatively different, the qualitative
features are robust and the dominant growth wave vectors remain the
same, so in the experiment strong signatures of ferromagnetism should
be observed.

\section{Experimental observation}

\begin{figure}
 \centerline{\resizebox{0.95\linewidth}{!}{\includegraphics{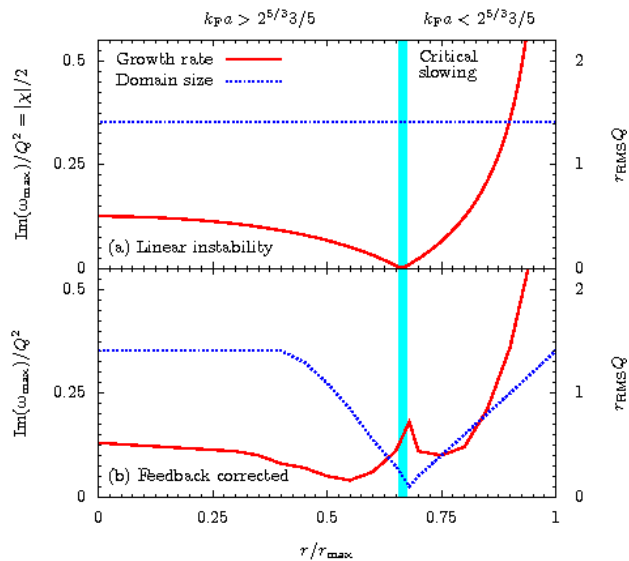}}}
 \caption{(Color online) The variation of growth rate (red solid,
   primary y-axis) and domain size (blue dashed, secondary y-axis)
   with radius within a harmonic well. (a) corresponds to linear
   response, and when (b) feedback corrections with a seed field of
   size $0.2Q^{2}/g$ are taken into account. The vertical cyan line
   denotes the radius corresponding to critical slowing and
   $r_{\text{max}}$ is the outer radius of the atom distribution.}
 \label{fig:TwistFerroTrap}
\end{figure}

Having studied the instability to a ferromagnetic state in a uniform
gas we now turn to consider the experimental ramifications of our
results. We will focus on two key questions: first the consequences
of the atomic gas being held within a realistic harmonic trapping
potential, and second we address the experimental signatures of our
predictions.

We shall treat the harmonic confinement within the local density
approximation.  On passing radially outwards from the center of the
harmonic well the local density and therefore the effective
interaction strength falls. We can therefore use the results of the
previous sections to map the rate at which the instability develops
and its characteristic wavelength as a function of the radius. These
results are shown in \figref{fig:TwistFerroTrap} for both the linear
analysis (a) and when the feedback due to scattering on the growing
mode is taken into account (b). Note significant modifications due to
feedback corrections. First near the radius which corresponds to
critical interactions there is a distinct maximum rather than a
vanishing growth rate. The enhanced growth rate is driven by a
scattering off the growing mode into large wave vectors. Therefore,
secondly the characteristic wavelength of the unstable modes dips
around that radius. In the linear analysis, by contrast the
characteristic wavelength of the unstable modes is everywhere
$2\pi\sqrt{2}/Q$.


\begin{figure}
 \centerline{\resizebox{0.95\linewidth}{!}{\includegraphics{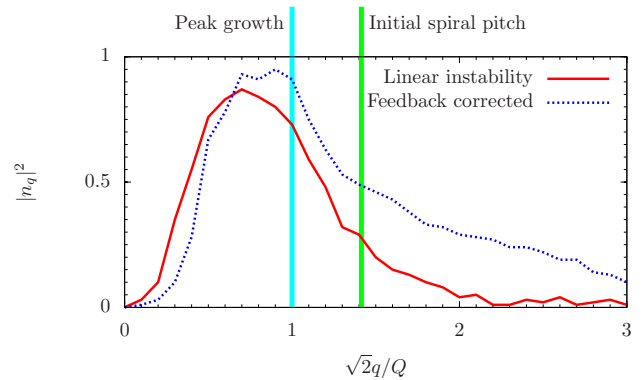}}}
 \caption{(Color online) The Fourier transform of the domain
   distribution within a harmonic well. Without magnetization feedback
   is shown by the red solid line, and the blue dashed curve is in the
   presence of magnetization feedback. The peak growth rate at
   $Q/\sqrt{2}$ is labelled by the vertical cyan line, whereas the
   maximum growth wave vector $Q$ is shown by the vertical green
   line.}
 \label{fig:TwistFerroFourier}
\end{figure}

We now turn to the question of how the distinct spatial structure of
the instability may be observed in an experiment.  The first approach
we look at is to employ differential in-situ phase-contrast imaging, a
method that has already been used on the ferromagnetic cold atom
gas~\cite{Jo09}. This measures the difference between the up and
down-spin populations integrated along vertical columns through the
gas. If the procedure is repeated across the gas, a two-dimensional
map will be produced, which is governed by the pattern of the emergent
magnetization aligned normal to the imaging plane.  Once Fourier
transformed, the spectrum of this map will reveal the typical magnetic
domain size.  To simulate the expected experimental result we employ a
heuristic model of the gas. We divide the system into domains, with
position dependent size, determined by the wavelength of the dominant
growth mode at that radius. Each domain is then assigned either up or
down magnetization at random, and the resulting magnetization
structure is column integrated to obtain a two dimensional
magnetization image as in the experiment. The power spectrum of the
domain structure in the simulated image is shown in
\figref{fig:TwistFerroFourier}.  Without feedback corrections,
according to \eqnref{eqn:ContinuumBareDispersion} the dominant mode is
at $q=Q/\sqrt{2}$ irrespective of interaction strength. In the
numerical experiment a distinct peak exists at wave vectors around and
below $Q/\sqrt{2}$; this is because adjacent domains were not
necessarily misaligned thus increasing the effective length scale of
the observed domains. If feedback corrections are taken into account
then in the region of the trap corresponding to critical slowing
significantly smaller domains are formed. This is reflected in the
Fourier spectrum with reduced weight at small wave vectors, and
enhanced weight at larger wave vectors. By tuning with either the size
of the seed field or the duration of the experiment, the signal of the
contrasting behavior when feedback corrections are taken into account
could help investigators to identify the ferromagnetic transition.

One criticism of the recent experiment~\cite{Jo09} which
reported the first signs of ferromagnetism is that the domains, if
present, were too small to image. In the new experimental protocol the
pitch length of the initial spin spiral sets the size of the domains
formed. However, if this length scale remains below the resolution of
the experiment then a statistical analysis of the in-situ
phase-contrast imaging could still provide an estimate of the domain
size.  If the orientation of adjacent domains within a column is
independent, then measurements over adjacent columns will give an
estimate of the variance of magnetization in each column. The
uncertainty in the net magnetization should vary as $sN/4\sqrt{n}$,
where $s$ is the spin per atom, $N$ the number of atoms imaged, and
$n$ the number of domains. A larger number of domains will result in a
smaller variance in the net magnetization, thus allowing the total
number of domains to be estimated.

One final experimental probe is spin-dependent Bragg spectroscopy. This
could track the decay of the planar spin spiral and the reducing
signature at wave vector $\vec{P}$. Furthermore, using a variable
wavelength optical lattice potential to couple asymmetrically to the
spin degrees of freedom, the collective mode response could
be studied through dynamical fluctuations of the cloud spatial
distribution as a function of wavelength, laser amplitude, and
detuning.

\section{Conclusions}

We have studied the time evolution of a spiraling spin texture,
prepared by a magnetic field gradient in an interacting degenerate
Fermi gas.  The linearized dynamics of the magnetization shows an
instability, which develops canting out of the spiral plane, with the
maximally unstable mode at a wavevector $1/\sqrt{2}$ that of the
original spiral. The instability grows exponentially in time with a
characteristic time scale that vanishes close to the Stoner transition
point. To linear order, the dynamics is the same on the
ferromagnetic and paramagnetic sides of the critical slowing down.

Interestingly, however, we find that near the critical point, non
linear effects in the dynamics become important. Specifically,
scattering of collective excitations on the exponentially growing
unstable mode acts to renormalize the spectrum, shifting the
instability to larger wave-vectors on approaching the original
critical point. Moreover the point of critical slowing is eliminated
and the new branch of excitations at high wave-vectors allows a clear
distinction between the ferromagnetic and paramegnetic regimes.

Finally, we studied the experimental signatures of the unstable
dynamical modes in a realistic trap confinement. Using the calculated
dispersion of unstable modes we obtained the spatial distribution of
spin-domain sizes in the trap, which may be observed with phase
contrast imaging. The most dramatic signature of the new branch of
instabilities induced by the non linear feedback effect is a collapse
of the domain size at a particular critical radius in the trap.

We thank Eugene Demler, Ben Simons, and especially
Gyu-Boong Jo, Wolfgang Ketterle, and Joseph Thywissen for useful
discussions. GJC acknowledges the financial support of the Royal
Commission for the Exhibition of 1851 and the Kreitman Foundation. EA
was supported in part by ISF, US-Israel BSF, and the Minerva
foundation.

\end{document}